**SEARCHING FOR COST OPTIMIZED INTERSTELLAR BEACONS**

Gregory Benford
Physics and Astronomy Dept.
University of California Irvine, Irvine CA 92612 USA

James Benford
Microwave Sciences, Inc. USA
Lafayette CA 94549 USA
Phone: 925-283-8454 Fax: 925-283-8487
jbenford@earthlink.net

Dominic Benford
Observational Cosmology Lab, NASA/Goddard Space Flight Center, Greenbelt MD
20771 USA

Running title:  Searching for Cost Optimized Interstellar Beacons

What would SETI Beacon transmitters be like if built by civilizations with a variety of motivations, but who cared about cost? We studied in a companion paper how, for fixed power density in the far field, we could build a cost-optimum interstellar Beacon system. Here we consider, if someone like us were to produce a Beacon, how should we look for it?  High-power transmitters might be built for wide variety of motives other than two-way communication; Beacons built to be seen over thousands of light years are such. Altruistic Beacon builders will have to contend with other altruistic causes, just as humans do, so may select for economy of effort. Cost, spectral lines near 1 GHz and interstellar scintillation favor radiating frequencies substantially above the classic "water hole." Therefore the transmission strategy for a distant, cost-conscious Beacon will be a rapid scan of the galactic plane, to cover the angular space.  Such pulses will be infrequent events for the receiver.  Such Beacons built by distant advanced, wealthy societies will have very different characteristics from what SETI researchers seek.  Future searches should pay special attention to areas along the galactic disk where SETI searches have seen coherent signals that have not recurred on the limited listening time intervals we have used.  We will need to wait for recurring events that may arrive in intermittent bursts. Several new SETI search strategies emerge from these ideas.  We propose a new test for SETI Beacons, based on the Life Plane hypotheses.

# 1. Introduction

As the 50[th] anniversary of the first SETI observation (Project Ozma) approaches in 2010, we should study the underlying conventional wisdom behind the search. With no detections in the near-zone search (~500 stars within ~ few hundred light years), SETI is nearing the rough limit in which optical data on candidate stars is useful (Brin, 1983). Beyond ~1000 light years, interstellar obscuration makes identifying telltale biological features such as an ozone spectral line difficult. SETI ranges > 1000 light years require an Effective Isotropic Radiated Power (EIRP) >$10^{17}$ W, so that the broadcaster enters the domain not of targeted radiators, but of Beacons. Our companion paper, "Messaging With Cost Optimized Interstellar Beacons" (Benford *et al.,* 2010), addresses how we would build such a Beacon.

The traditional targeted SETI strategy had much to recommend it. The background noise minimum in the "water hole" region near 1 GHz seemed plausible, as did the assumption that the altruistic radiator would beam forth steady, targeted signals of very narrow bandwidth, to make detection by us easy.

Recent developments have lessened the power of these early views.

## 1.1 The Galactic Habitable Zone

There is a growing sentiment within the astrobiological community that we are not typical members of the suite of galactic civilizations because we live among the outer regions of a Galactic Habitable Zone (Kasting *et al.*, 1993; Trimble, 1997; Gonzales *et al.*, 2001). In papers such as Lineweaver (2001), and in popularizations such as *Rare Earth* (Ward and Brownlee, 2000) a view emerged that stresses the difficulties facing intelligent life in our galaxy; intelligence may be as rare as ducks in a desert. Lineweaver argues that early, intense star formation toward the inner Galaxy provided the heavy elements necessary for life, but the supernova frequency remained dangerously high there for several billion years. Later, stars orbiting between the crowded inner bulge and the barren outer Galaxy were born into a habitable zone, starting about 8 Gy ago. The habitable zone expanded with time as metalicity (driven by supernovas) spread outward in the Galaxy and the supernovae rate decreased. They argued that ~ 75% of the stars that harbor complex life in the Galaxy are older than the Sun and that their average age is ~ 1 Gy older than the Sun.

This implies that most advanced societies should lie much farther inward toward the galactic center, at distances > 1000 light years. Listening to relatively local star systems (SETI's primary strategy for decades) misses most of the possible civilizations. (For a contrary view, see Vukotic and Cirkovic, 2007.). It doesn't follow the Galactic Center Strategy we describe below. Such distances imply for the Beacon builder rather different motives than for nearby emitters.

# 2. Cost Of Alien Beacons

We recently studied how, for fixed power density in the far field, we could build a cost-optimum interstellar Beacon system on Earth (Benford J. *et al.,* 2010). Here we consider, if someone else were to produce a Beacon, how should we look for it? That is, we apply our arguments, based on terrestrial cost-determined design techniques, to alien societies.  But are there any galactic social universals?  Possibly, there are none in common between aliens and us, so why should any arguments inspiring SETI have any weight?  SETI assumes the opposite—that we can have similar motives.

 Are aliens unknowable, and beyond economic arguments?  We'll call this the Altruistic Alien argument -- that aliens of great ability, near-infinite resources and benign intent will transmit to us without taking any consideration to the cost (which would be high in our terms).  This argument is seldom directly expressed.

But this argument meets a conceptual danger: If Altruistic Aliens have great resources, they would find it easy to make themselves apparent in our night sky.  If so, where are they?  We now know that within a range of ~ 400 light years they do not make themselves obvious. So, Beacons are necessary beyond several 1000 light years. No conversations occur over such scales; transmissions are announcements or memorials, not letters.

We assume that if they are social beings interested in a SETI conversation (Hetesi and Regály, 2006) or passing on their heritage, they will know about tradeoffs between social goods, and thus, in whatever guise it takes, *cost*.  But what if we suppose, for example, that aliens have very low cost labor, i.e., slaves?  With a finite number of slaves, you can use them to do a finite number of tasks.  And so you pick and choose by assigning value to the tasks, balancing the equivalent value of the labor used to prosecute those tasks.  So choices are still made on the basis of available labor.  The only case where labor has no value is where labor has no limit.  That might be if aliens may live forever or have limitless armies of self-replicating automata, but such labor costs something, because resources, materials and energy, are not free.

Our point is that *all* SETI search strategies must assume something about the Beacon builder, and that cost may drive some alien attempts at interstellar communication.

## 2.2. Beacon–builder Motives

Through most of its history SETI has assumed a high-minded search for other lifeforms.  But other motives are possible.

What could motivate a Beacon builder?  Here we can only reason from our own historical experience. Other possible high intelligences on Earth (whales, dolphins, chimpanzees) do not have significant tool use, so they do not build lasting monuments. Sending messages over millennia or more connects with our own cultures.  Human history suggests (Benford G., 1999) that there are two major categories of long-term messages that finite, mortal beings send across vast time scales:

- *Kilroy Was Here* These can be signatures verging on graffiti. Names chiseled into walls have survived from ancient times. More recently, we sent compact disks on interplanetary probes, often bearing people's names and short messages that can endure for millennia.
- *High Church* These are designed for durability, to convey the culture's highest achievements. The essential message is *this was the best we did; remember it*.

A society that is stable over thousands of years may invest resources in either of these paths. The human prospect has advanced enormously in only a few centuries; the lifespan in the advanced societies has risen by 50% in each of the last two centuries. Living longer, we contemplate longer legacies. Time capsules and ever-proliferating monuments testify to our urge to leave behind tributes or works in concrete ways (sometimes literally). The urge to propagate culture quite probably will be a universal aspect of intelligent, technological, mortal species (Minsky, 1985).

Thinking broadly, high-power transmitters might be built for wide variety of goals other than two-way communication driven by curiosity. For example:

- *The Funeral Pyre:* A civilization near the end of its life announces its existence.
- *Ozymandias:* Here the motivation is sheer pride; the Beacon announces the existence of a high civilization, even though it may be extinct, and the Beacon tended by robots. This recalls the classic Percy Bysshe Shelly lines,
  > *And on the pedestal these words appear:*
  > *'My name is Ozymandias, King of Kings;*
  > *Look on my works, Ye Mighty, and despair!'*
  > *Nothing beside remains. Round the decay*
  > *of that colossal wreck, boundless and bare,*
  > *The lone and level sands stretch far away.*
- *Help!* Quite possibly societies that plan over time scales ~1000 years will foresee physical problems and wish to discover if others have surmounted them. An example is a civilization whose star is warming (as ours is), which may wish to move their planet outward with gravitational tugs. Many others are possible.
- *Leakage Radiation*: These are unintentional, much like objects left accidentally in ancient sites and uncovered long after. They do carry messages, even if inadvertent: technological fingerprints. These can be not merely radio and television broadcasts radiating isotropically, which are fairly weak, but deep space radar and beaming of energy over solar system distances. This includes "industrial" spaceship launchers, beam-driven sails, "planetary defense" radars scanning for killer asteroids, and cosmic power beaming driving interstellar starships with beams of lasers, millimeter or microwaves. There are many ideas about such uses already in the literature (Benford & Benford, 2006).
- *Join Us*: Religion may be a galactic commonplace; after all, it is here. Seeking converts is common, too, and electromagnetic preaching fits a frequent meme.

Whatever the Beacon builders' motives, we should periodically reassess our SETI assumptions in light of how our own microwave emitting technologies develop. Since the early SETI era of the 1960s, microwave emission powers have risen orders of magnitude and new technologies have altered our ways of emitting very powerful signals. Given Beacon ranges > 1000 ly, EIRPs >$10^{16}$ W are needed. These high powers suggest that all possible motivations will succumb to economics, however. Is cost/benefit analysis arguably universal?

## 2.2. Parsimony and Beacons

Traditional SETI research takes the point of view of *receivers*, not *transmitters*. This neglects the implications for what signals should look like *in general*, and especially the high emitting costs, which a receiver does not pay.

We shall assume, like conventional SETI, that microwaves are simpler for planetary societies, since they can easily outshine their star in microwaves. Microwaves are probably better for Beacons (Tarter, 2001).

Whatever the life form, evolution will select for economy of resources. This is an established principle in evolutionary theory (Williams, 1966). Further, Minsky (1985) argues that a general feature of intelligence is that it will select for economy of effort, whatever the life form. Tullock (1994) argues that social specie evolve to an equilibrium in which each species unconsciously carries out "environmental coordination" which can follow rules like those of a market, especially among plants. He gives many such examples. Economics will matter.

A SETI broadcaster will face competing claims on resources. Some will come from direct economic competition. Standing outside this, SETI beaming will be essentially altruistic, since replies will take centuries if not millennia, or else are not even an issue. SETI need not tax an advanced society's resources. The power demands in our companion paper are for average powers $\leq$GW, far less than the 17 TW now produced globally (Hoffert *et al.*, 2002)

But even altruistic Beacon builders will have to contend with other competing altruistic causes, just as humans do (Lemarchand and Lomberg, 1996). They will confront arguments that the response time for SETI is millennia, and that anyway, advanced societies leak plenty of microwaves etc. into deep space already. We take up these issues below. It seems clear that for a Beacon builder, only by minimizing cost/benefit will their effort succeed. This is parsimony, meaning 'less is better' a concept of frugality, economy. Philosophers use this term for Occam's Razor, but here we mean the press of economic demands in any society that contemplates long term projects like SETI. On Earth, advocates of METI (Messaging to Extraterrestrial Intelligence) will also face economic constraints (Benford *et al.*, 2010).

Note that parsimony directly contradicts the Altruistic Alien Argument that the Beacon builders will be vastly wealthy and make everything easy for us. An omnidirectional Beacon, radiating at the entire galactic plane, for example, would have to be enormously powerful and expensive, and so not be parsimonious. One of the SETI founders, B. Oliver, calculated a cost minimization for both sender and receiver together, but Oliver's sender and receiver were not part of the same economic system, and indeed do not know each other, so there is no reason for cost to be minimized between them.

Parsimony has implications for SETI. For transmitting time $\tau$, receiver detectability scales as $\tau^{1/2}$. But at constant power, transmitter cost increases as $\tau$, so short pulses are economically smart (cheaper) for the transmitting society. A one-second pulse sent every 10 minutes to 600 targets would be 1/600 as expensive per target, yet only ~1/25 times harder to detect. Interstellar scintillation limits the pulse time to $> 10^{-6}$ sec, which is within the range of all existing HPM devices. Such pings would have small information content, attracting attention to weaker, high content messages. This general scaling seems clear, but of course the crucial issue is what frequencies and time scales are best.

Finally, even if Earth economics generally works similarly in other technological societies, why should it apply to their transmitting Beacons? Even on Earth, larger goals often override economic dictates, such as military security, aesthetics, religion, etc. But two aspects of SETI undermine this intuition:

1. SETI assumes long time scales for sender and receiver. But while cultural passions can set goals, economics determines how they get done. Many momentary, spectacular projects such the pyramids of Egypt lasted only a century or two, then met economic limits. The Taj Mahal so taxed its province that the second, black Taj was never built. The grand cathedrals of medieval Europe suffered cost constraints and, to avoid swamping local economies, so took several centuries of large effort. Passion is temporary, while costs constrain long-term projects.

2. The optimum cost strategy leads directly to a remarkable cost insensitivity to the details of economic scaling. *The ratio of costs for antenna area and system power depends on only the ratio of exponents (Benford et al., 2010, eq. 13)—and not on the details of the technology. That ratio varies on Earth by only a factor of two. Both these costs may well be related principally to labor cost; if so it cancels out. This means fashions in underlying technology will matter little, and our experience perhaps robustly represents that of other technological societies.*

## 3. Consequences for SETI Strategy

If there are cost-optimized Beacons, we argue they can be found by steady searches that watch the galactic plane for times on the scale of years. Of course SETI literature abounds with consideration of the tradeoffs of search strategy (range vs. EIRP vs. pulse vs. CW vs. polarization vs. frequency vs. beamwidth vs. integration time vs. modulation types vs. targeted vs. all-sky vs. Milky Way). But in practice, search dwell times are a few seconds in surveys and 100 to 200 seconds for targeted searches (Tarter, 2001). Optical searches usually run to minutes. And integration times are long, of order 100 sec, so short pulses will be integrated out. Recent discoveries of transient signals (Hyman, 2007) have sparked renewed interest in shorter observation times. With such searches becoming more common, detection of short-pulse Beacons becomes more likely.

*3.1 Bandwidth and Frequency*

At distances >1000 light years Doppler adjustment to offset relative motions, as nearby SETI searches do, becomes pointless; with many stars in the field of view, none is especially addressed. Further, distortion of signals from >1000 light years arises from interstellar scintillation. Such "twinkling" of the signal comes from both the dispersion of differing frequencies, and delays in arrival time for pulses moving along slightly different pathways, due to refraction. Temporal broadening probably would limit bandwidth to >1 MHz, as we know from the broadening of pulsar signals.

*3.2 Have We Seen Beacons?*

Behind conventional SETI methods lies the assumption that altruistic beaming societies will send persistent signals. In searches to date, confirmation attempts, when the observer looks back at a target, in practice usually occur days later (or for data found in the SETI At Home program, years). Such surveys have little chance of seeing cost-optimized Beacons.

For example, Shostak (2008) argues that a Beacon would need to dwell on $\sim 10^5$ stars to have a reasonable choice of getting through to a working receiver. He bases this on a straw man calculation that assumes 10,000 transmitting societies in the galaxy with an average lifetime of 10,000 years, implying a mean separation of 1000 light years. His crucial assumption is that to give the receiver a chance to confirm, the Beacon should remain on target for a week, a confirmation time based on experience, and typical in contemporary sky surveys (Tarter, 2001). Thus, it would take 2000 years to dwell on the $10^5$ stars. A group of $10^5$ stars is, for example, a large sample of the nearby stars that would be thought of as likely being productive targets. A better strategy would be to dwell on many stars for the emitting time of the high power device, perhaps a minute or even less than a second. Then repeat the dwell pattern, perhaps after a week. An observer would then have a chance to confirm the contact by looking back at likely events for times of days. Somewhere in that time the signal would reappear, perhaps a brief burst with coded pointers to a lower power, high content signal.

Why would one-second dwells make sense? Certainly enough information can be conveyed in a second to be a Beacon and definitively non-natural. On our nomograph (Fig.11, Benford *et al.,* 2010), taking a typical (for Earth technology) $10^{-4}$ radian beam, a second dwell time means $F \sim 10^{-4}$, so, from eq. 6 repeat time is only a day. If one chooses to argue that it would be better to dwell for, say, an hour, then one revisits every ten years.

What does a cost-optimized, pulsed, broadband, narrow focus Beacon emission imply? Distant, cost-optimized Beacons will appear for much less time than as assumed in conventional SETI. Earlier searches have seen pulsed intermittent signals resembling what we (in this paper) think Beacons may be like, and may provide useful clues. We should observe the spots in the sky seen in previous work for hints of such activity but over yearlong periods. Natural radio source variability does extend down at least to the ~month timescale, and so intelligent civilizations will perhaps be looking for signals relevant to month-long revisit times. Perhaps newer search methods, directed at short transient signals, will be more likely to see the Beacons we have described (Cordes *et al.*, 2007,Siemion *et al.*, 2008, Lazio *et al.*, 2009).

The opposite strategy is, if you really wanted to be heard by a known -- or suspected -- extraterrestrial civilization, you could spend a large fraction of your time on them, dwelling for months. The nomograph indicates that this strategy works well for revisit times of order a year. (In our nomograph, this is the 'continuous beaming' quadrant.) With resources and technology appropriate to our current capabilities, our optimized Beacon beamwidth would be such that the number of civilizations we could look at would be a few per year. An active and stable SETI program could provide good new targets at such a rate, based on the occurrence of interesting signals seen over recent decades. This strategy is the inverse of the above: the targets are not random, but are preselected based on existing observational information. Potentially biased, yes, but based on some knowledge -- and hence more likely to be worth the investment of time broadcasting to them.

A provocative example is Sullivan, *et al.* (1997). This survey lasted about 2.5 hours, with 190, 1.2 minute integrations. They "recorded intriguing, non-repeatable, narrowband signals, apparently not of manmade origin and with some degree of concentration toward the galactic plane…" Similar searches also saw one-time signals, not repeated (Shostak &Tarter, 1985; Gray & Marvel, 2001, Gray, 1994; 199 Tarter, 2001). These searches had slow times to revisit or reconfirm, often days (Tarter, 2001). Overall, few searches lasted more than hour, with lagging confirmation checks (Horowitz & Sagan, 1993).

Another striking example is the "WOW" signal seen at the Ohio SETI site. The check-back time was fairly long, and subsequent studies observed for short times. However, the system rejected signals greater than 10 kHz wide at 1.5 GHz, i.e., no more than $\Delta f / f \sim 10^{-5}$. A broadband HPM source would usually have $\Delta f / f \sim 10^{-3}$ (Benford, *et al*, 2010, Appendix B), so the Ohio search may have simply missed events. Further, the total time spent by all later searches of the WOW signal site, directly toward galactic center, is about 1% of a year. This fact illuminates the constraints that a Galactic Center Search Strategy imposes: a yearlong campaign will require more effort than SETI has enjoyed over the last half century.

These two are examples among many set forth in the SETI Institute's database (http://www.seti.org/Page.aspx?pid=689). That set of striking but later unconfirmed detections should be the basis of a persistent search. There is increasing interest in transient radio phenomena, and Beacons such as we describe are within the parameer space of such sources (Cordes, 2007).

### 3.3 Is GCRT J17445-3009 a Beacon?

As an example of using cost optimized Beacon analysis for SETI purposes, consider in detail the puzzling transient bursting radio source, GCRT J17445-3009, which has extremely unusual properties. It was discovered in 2002 in the direction of the Galactic Center (1.25° south of GC) at 330 MHz in a VLA observation and subsequently reobserved in 2003 and 2004 in GMRT observations (Hyman, *et al*., 2005, 2006, 2007). It is a pulsed coherent source, with the 'burst' lasting as much as 10 minutes, with 77-minute period. Averaged over all observations, Hyman et al. give a duty cycle of 7% (1/14), although since some observations may have missed part of bursts, the duty cycle might be as high as 13%.

What is it? Candidate explanations include masers, flare stars, extrasolar planets, periodic or precessing radio pulsars, double neutron star binary pulsars, white dwarf pulsars (J. Echevarría, *et. al*., 2008) and ultracool brown dwarfs. Nothing fits well, and observations continue. Could this source be a Beacon?

We take this case because, unlike the spotty potential SETI sightings like the "Wow" event, GCRT J17445-3009 has been seen repeatedly. Any true SETI source will face similar scrutiny and analysis.

For Messaging to Extraterrestrial Intelligence (METI), experience shows an optimum tradeoff, depending on transmission frequency and on antenna size and power (Benford *et al*, 2010). This emerges from minimizing the cost of producing a desired effective isotropic radiated power (EIRP), which in turn determines the maximum range of detectability of a transmitted signal. Costs of pulsed cost-efficient transmitters were estimated from these relations using current cost parameters ($/W, $/m²) as a basis. The result is that galactic-scale Beacons demand effective isotropic radiated power $>10^{17}$ W, emitted powers are >1 GW, with antenna areas ~ km². Thrifty Beacon systems would be large and costly, have narrow 'searchlight' beams and short 'dwell times' when the Beacon would be seen by an alien observer at target areas in the sky. They may revisit an area infrequently. The natural corridor to broadcast is along the galactic spiral's radius or along the spiral galactic arm we are in.

Applying this approach to CRT J17445-3009, the observed flux density is about 1 Jansky at 330 MHz and the bandwidth is at least 30 MHz (Hyman *et al*, 2007). This implies a minimum power density of S=$3 \cdot 10^{-19}$ W/m$^2$. As the source is very close in angle to the Galactic Center, we begin by assuming that GCRT J17445-3009 is a Beacon located there, $R_{GC}$=26,000 ly from Earth.

The power density S of a Beacon at range R is determined by W, the effective isotropic radiated power (EIRP), the product of radiated peak power P and aperture gain G,

$$W = PG$$
$$S = \frac{W}{4\pi R^2} \tag{1}$$

and gain is given by area and wavelength:

$$G = \frac{4\pi\varepsilon A}{\lambda^2} = \frac{4\pi\varepsilon A}{c^2}f^2 = kAf^2$$
$$W = kPAf^2 \tag{2}$$
$$k = \frac{4\pi\varepsilon}{c^2}$$

Here e is aperture efficiency (this includes factors such as phase, polarization and array fill efficiency) and we have collected constants into the factor k. From Eq. 1, W=$2.3 \cdot 10^{23}$ W, a large Beacon.

Such a high power radiator would be extremely expensive, so we use our analysis of Beacon cost (Benford et al., 2010), which shows that the optimum power and antenna area for producing the power density S is

$$P^{opt} = \sqrt{\frac{aW}{pkf^2}}$$
$$A^{opt} = \sqrt{\frac{pW}{akf^2}} \tag{3}$$

Here we minimize the cost by assuming linear scaling dependence of cost on the peak power and antenna area, a well-established method in industry. Coefficients describe the dependence of cost on area a ($/m$^2$), which includes cost of the antenna, its supports and sub-systems for pointing and tracking and phase control, and microwave power p ($/W) which includes cost of the source, power supply, cooling equipment and prime power:

$$C_A = aA$$
$$C_S = pP$$
$$C_C = aA + pP \qquad (4)$$

Using rough estimates of costs on Earth (1000 \$/m$^2$, 1\$/W), antenna efficiency 60%, one can estimate features of Beacons. From Eq. 3 and the observed 330 MHz frequency, $P^{opt}$=3.2 TW, $A^{opt}$=9500 km$^2$, and the optimum antenna diameter $D^{opt}$=110 km. The minimized cost is $C^{opt}$=19 T\$ for 3\$/W microwave tubes, 10 k\$/m$^2$ antennas. For solid-state costs, we use the scaled estimates (Scheffer, 2005, Table 3) and find the cost is about the same, about 20 T\$. So they could use either of the technologies available to us, longer wavelengths (solid state) or shorter (electron beam tubes).

The angular width θ of the optimized Beacon beam is set by the antenna area and frequency:

$$\theta = c\left[\frac{ak}{f^2pW}\right]^{1/4} \qquad (5)$$

For this case, θ=9·10$^{-6}$ sr. The *dwell time* when the Beacon beam falls on the receiver of a listener $\tau_d$ is related to the *revisit time* or cycle time $\tau_r$ that the Beacon with optimal emitting angle $\theta_o$ takes to broadcast across a segment of the galactic plane $A_G$. F is the fraction of the sky $A_G$ covers, as seen from the Beacon:

$$\frac{\tau_d}{\tau_r} = \frac{A(\Theta_o)}{A_G} = \frac{\pi\theta^2}{4A_G} = \frac{\theta^2}{16F} \qquad (6)$$

For the GCRT J17445-3009 observations, dwell time is 10 minutes, revisit time is 77 minutes. The fraction F of the sky the scanning Beacon covers is 4x10$^{-11}$. The beam spot at Earth is θR$_{GC}$= 0.24 ly. At our local star density of 0.2 star/ly$^3$, the beam illuminates at most one star at our distance. The beam in the 26,000 ly path to Earth illuminates ~100 stars (Sullivan *et al*, 1984).

If the Beacon, with 7% duty cycle, is painting 14 such 0.24 ly areas, then repeats 77 minutes later, then the total illuminated area is a circle 1 ly in diameter, and the beam illuminates ~1 star in a 1 ly diameter sphere.

If the Beacon is closer, it is smaller and cheaper: For R=1000 ly, W=3.4·10$^{20}$ W, still a large Beacon, $P^{opt}$=4.7 GW, $A^{opt}$=366 km$^2$, $D^{opt}$=22 km. The minimized cost is $C^{opt}$=730 B\$, θ=4.7·10$^{-5}$ rad, the beam spot is θR= 4.7·10$^{-5}$ ly. At our local star density of 0.2 star/ly$^3$, the beam illuminates at most one star at our distance.

From the above, GCRT J17445-3009 probably isn't a Beacon. But if it is a cost-optimized Beacon, it must be targeting because the field of stars covered is so small. It's trying to talk to our star system. Perhaps the Beacon is in fact much closer and targets our region in part because they have detected signs of life in Earths atmosphere such as out-of equilibrium chemistry (oxygen, ozone). Or the Beacon can have a hierarchy of timescales, illuminating one part of its sky intensively, then moving on to another promising region, returning to us years later.

Another possibility is that a cost-optimized Beacon is engaged in communication along the natural radial corridor from Galactic Center, knowing that astronomical races will study the Center preferentially (Fig. 1). We just happen to be in their beam, having intercepted an interstellar communication link. In this case, we should:

1) Stare at the direction of GCRT J17445-3009 at higher frequencies as both cost optimization and higher information-carrying ability argue. Another information-bearing signal could be at the optimum high frequencies, ~10 GHz. A temporal analysis should be conducted to search for structure in the bursts, since measurements to date have not looked for any message content.

2) look in the opposite direction, 180 degrees from the Center, to see if there's another beam communicating toward GCRT J17445-3009.

Why would a Beacon be at such a low frequency as 0.33 GHz, which is not cost optimum? Those seeking to reach emergent tech societies may want to go to lower frequencies, because the seemingly natural evolution of electromagnetic technology is from long wavelengths, then at later times to shorter ones. Long wavelength engineering is easier, as Marconi first showed; radio astronomy started at low frequencies. The same technology path means that the first radio astronomy observations will be at low frequencies (as for Jansky) and detect the strongest sources. If own experience is typical, it supports the above assertion.

### 3.4 A Galactic Center Search Strategy

To see Beacons as we envision them, we should search in the plane of the spiral disk. From Earth, 90% of the galaxy's stars lie within 9% of the sky's area, in the plane and hub of the galaxy. This suggests a limited sky survey. We will need to be patient and wait for recurring events that may arrive in intermittent bursts. Special attention should be paid to areas along the Galactic Disk where SETI searches have seen coherent signals that are non-recurring on their limited listening time intervals. Since most stars lie close to the galactic plane, as viewed from Earth, occasional pulses at small angles from that plane should have priority.

Whatever forms might dwell further in from us toward the center, they must know the basic symmetry of the spiral. This suggests the natural corridor for communication is along the spiral's radius from Galactic Center or toward it, a simple direction known to everyone. (A radius is better than aiming along a spiral arm, since the arm curves away from any straight-line view of view. On the other hand, along our nearby spiral arms the stars are roughly the age of ours.) This avenue maximizes the number of stars within a telescope's view, especially by staring at the galactic hub. Thus, a Beacon near the center should at least broadcast outward in both directions, while societies at the far reaches may save half their cost by not emitting outward, since there is much less chance of advanced societies there. Radiating into the full disk takes far more time and power, so beams may only occasionally visit any sector of the radial plane. We listeners fairly far out (and fairly young) should look inward, within a narrow angle (~ 10 degrees) toward the constellation Sagittarius. Listening outward seems less efficient, since fewer life sites lie that way.

Life sites like ours will also know two rough time scales—a year and a day, from constraints on planetary habitable zones and biosphere mechanics. Observing every day over a year span might have a better chance of seeing intermittent bursts that revisit our part of the sky on a yearly time scale. To lower costs and have the best viewing range, sites near the equator seem optimal. The Indian GMRT group observes in meter wavelengths, up to 1.5 GHz, can see the galactic center year-round, and is well placed in a low-noise area. The GMRT cannot observe at the higher frequencies we advocate. Parkes is better situated at latitude -33 where the galactic center passes nearly overhead, and can see high frequencies.

*3.5 A Life Plane Strategy*

The 2004 discovery (Rohde and Meuller, 2005), that marine life biodiversity follows a 62 million year cycle, suggests that a Life Plane may exist in our galaxy. To explain this observation Medvedev and Melott propose that the cycle arises from modulation of cosmic ray flux by our sun's vertical oscillation (~62 My period) above and below the galactic plane (Medvedev and Melott, 2007). They argue that the galaxy's bow shock, formed as it moves northward toward the Virgo cluster, enhances cosmic ray flux when our star is ~ 250 light years north of the galactic plane. The galactic halo/wind/termination shock ram pressure due to the galactic motion accelerates particles, amplifying the flux by a factor ~ 4.6 at the northern peak of our oscillation above the galactic plane, damaging the biosphere.

If even higher fluxes (and thus larger oscillation heights) suppress advanced life forms, this may define a plane near the center of the galactic 1000 light-year-thick disk that favors life, including intelligent life. Within this volume, stars oscillating vertically within perhaps < 500 light years will be favorable for civilizations, so Beacons will cluster about the plane, as well as be targets for a Beacon. Interestingly, the highest power transient sources observed by Horowitz & Sagan also lie close to the galactic plane.

If these Life Plane ideas are true, they could influence a Beacon builder's strategy. For example, confining the Beacon to the Life Plane simplifies the emitting pattern. (Note that our 230 ly amplitude oscillation about the plane is a small excursion. The scale length for the fall of star density going away from the plane is 650 ly. So the angle for our system seen from the galactic center is $\sim 10^{-2}$ radians.) Beacons with galactic ranges will be of narrow angle. The number of separate shots needed to illuminate the cylindrical surface at distance R and of height h(R), given an emitting angle $\theta$, is

$$N_h = \frac{2\pi R h(R)}{\pi\left(\dfrac{R\theta}{2}\right)^2} = \frac{8h(R)}{R\theta^2}$$

$$h(R) = \frac{R_0 h_0}{R} \qquad\qquad (7)$$

$$N_h = \frac{8R_0 h_0}{R^2 \theta^2}$$

Here the distance $R_0$ and height $h_0$ are scaling distances in the galactic plane. A Beacon following the Life Plane strategy then can target into narrower layers as R increases. For example, if $h_0$=500 ly, $R_0$=1000 ly, at R=4000 ly, for $\theta$ =2.5 $10^{-4}$, $N_h$= 8 $10^6$. If the Beacon sends a short signal to all spot sizes within $\theta$ in an Earth day, each shot appears in the receiver's sky for $10^{-2}$ sec. If the Beacon elects to illuminate only once a year, the signal appears for 4 sec. For longer ranges R, N drops, so pulses can last longer. The Life Plane strategy may be optimal, with less time spent at higher angles from the plane.

If life is clustered near the galactic plane, and if alien Beacons are following this strategy, then Beacons will be seen to cluster near the plane. If Medvedev and Melott are right, a *new test for SETI Beacons* is possible: Compare the distribution of the observations described in section 3.2 to the distribution of stars about the plane. If they differ, and in particular, if they are more tightly clustered toward the plane than all stars, it could be indirect evidence that some of the observations are of Beacons of the type we describe.

*3.6 Transit Targeted Search Strategy*

Nussinov (2009) suggested that inhabitants of star systems that lie close to the plane of the Earth's orbit around the sun could detect eclipsing by our annual transit across the face of the sun. That would tell them that Earth lies in a liquid water stable habitable zone (Corbet, 2003). Through spectroscopic analysis of our atmosphere, they might see traces of chemistry indicating that Earth likely bears life, particularly the prominent ozone line.

The more numerous distant observers of eclipsing would be at the intersections of the solar ecliptic with the galactic plane (in Taurus and Sagittarius), which are inclined at 60 degrees. They could target our system with a transmitting Beacon, so that is a preferred direction for us to search. The fraction of the sky that intersection fills is F ~$10^{-4}$. If thwy were to use a microwave Beacon, from Eq. 6 the dwell time would increase, the revisit time substantially shorten, or some combination of both. We can get a rough idea of such a Beacon in the Table of Beacon examples in our METI paper (Benford *et al.,* 2010), the short-pulse 1,000 ly-range Beacon could appear in our sky for 35 seconds three times a day, thus would be readably observable.

Targeted lasers would be the transmitters of choice, although they would not penetrate atmospheres of planets or hot-Jupiter moons with substantial atmospheres (Venus, Titan). Further, if they are used as an optical targeted laser Beacon, they would preferably emit at times whereby the signal arrives when the Earth is on the nearest (to the sender) half of its orbit, arriving on our night side. This gives a time-dependent term to an optical listening agenda.

*3.7 Implicit cooperation between Beacons and receivers*

Beacon builders who cost-minimize will expect receiving societies will have worked through the same calculations, as we now have, even if they do not build Beacons, as we have not. This implicit collaboration can insure that the receiver will invest in microwave antennas that maximize chances of detection, while minimizing their costs. This dictates a large receiving phased array antenna, optimized to capture brief signals that can occur at any time. If the receiver follows a radial strategy, preferentially looking toward galactic center, favoring the ~10 GHz region, the two parties maximize their chances of a connection. An alternative is a collection of smaller, perhaps privately operated, antennas around our planet, watching continuously for short-pulse Beacons.

Further, a brief SETI signal will plausibly carry information. If a candidate brief pulse carries a signal, this resolves the issue of whether it is a Beacon. Whatever coding strategy a signal uses, saving the phase information gathered by the receiver is essential. Radio astronomers typically measure frequency spectra, but keeping phase and frequency information should be added to our strategy for identifying SETI Beacons.

## 4. Conclusions

We conclude that SETI searches may have been looking for the wrong thing. SETI has largely sought signals at the lower end of the cost-optimum frequencies. They also may have taken needless care adjusting Doppler shifts, since broadband Beacons will need none. Searches have seen coherent signals that are non-recurring on their limited listening time intervals. Those searches may have seen Beacons, but could not verify them because they did not steadily observe for more than short periods.

We should reconsider SETI search strategies to enhance use of higher frequencies and make systematic scans of the entire galactic plane, with special attention to the galactic center. Searches for such signals might best be done in mid-latitude southern sites. We propose a new test for SETI Beacons, based on the Life Plane hypotheses. This requires steadily observing over periods of years.

We summarize the implications of these cost-minimized Beacon results as strategies:

1. Revisit the locations of the transient, powerful bursts seen in past surveys in a systematic way. Since we know these locations, a search every day or even more often would be inexpensive.

2. Scan the region pointing directly toward and away from the galactic center.

3. Scan the entire plane of the galaxy often throughout the year,

4. Since the highest nearby density of stars lies along the nearby galactic arm, listen in that direction for occasional, transient pulses.

5. Assume the Life Plane strategy of the Beacon builder. Observe a narrow range above and below the galactic plane daily.

**Acknowledgements**

We thank Paul Davies, Jill Tarter, Frank Drake, Paul Shuch and Seth Shostak for enlightening conversations and Jon Lomberg for the figure.

**Author Disclosure Statement**

No competing financial interests exist.

**Figure Legend**

Fig. 1 The recently observed source GCRT J17445-3009 can be a cost-optimized Beacon only if it is part of a narrowly directed radial interstellar communication link. (copyright Jon Lomberg, www.jonlomberg.com)

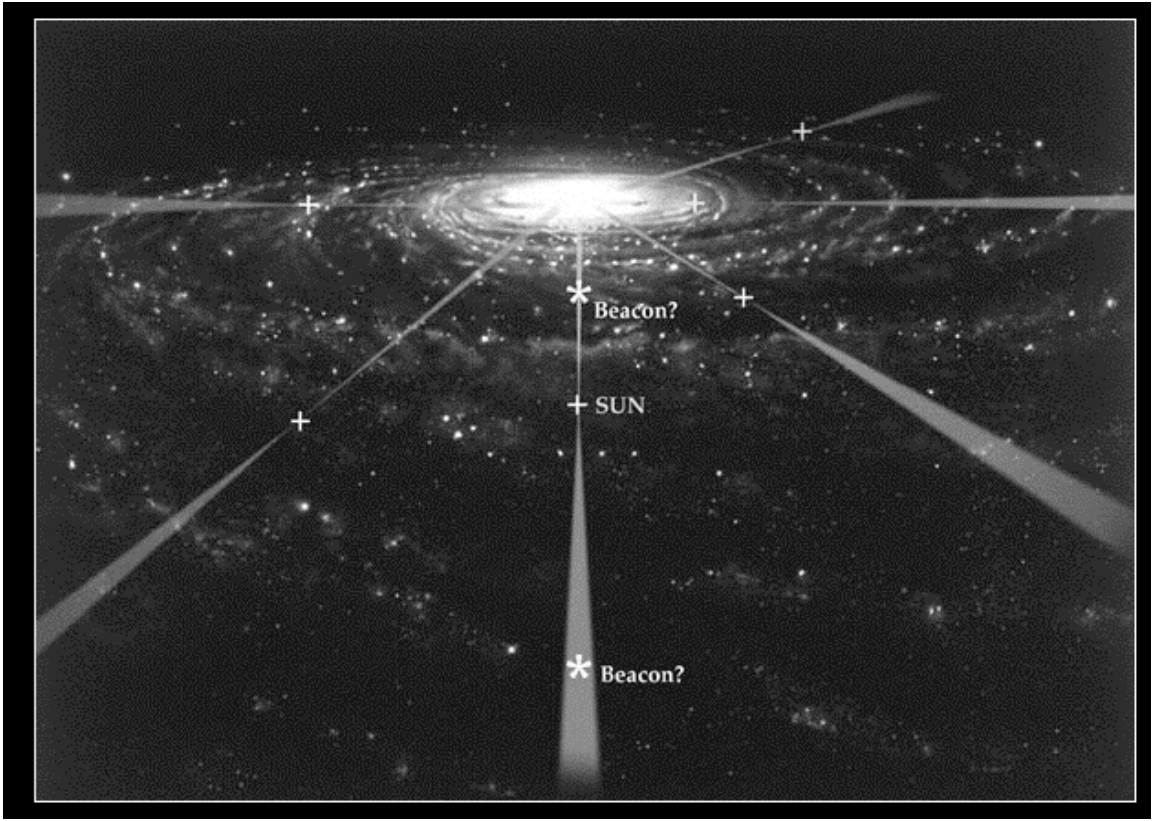